\documentclass{article}\usepackage{jkas2}
\runningauthor{ASAI, FUKUDA, \& MATSUMOTO}
\runningtitle{MHD SIMULATION OF SUBCLUMP}
\beginpage{1}
\endpage{5}

\begin{document}

\font\twelvei = cmmi10 scaled\magstep1 
       \font\teni = cmmi10 \font\seveni = cmmi7
\font\mbf = cmmib10 scaled\magstep1
       \font\mbfs = cmmib10 \font\mbfss = cmmib10 scaled 833
\font\msybf = cmbsy10 scaled\magstep1
       \font\msybfs = cmbsy10 \font\msybfss = cmbsy10 scaled 833
\textfont1 = \twelvei
       \scriptfont1 = \twelvei \scriptscriptfont1 = \teni
       \def\mit{\fam1 }
\textfont9 = \mbf
       \scriptfont9 = \mbfs \scriptscriptfont9 = \mbfss
       \def\bmit{\fam9 }
\textfont10 = \msybf
       \scriptfont10 = \msybfs \scriptscriptfont10 = \msybfss
       \def\bmsy{\fam10 }

\def\etal{{\it et al.~}}
\def\eg{{\it e.g.,~}}
\def\ie{{\it i.e.,~}}
\def\lsim{\raise0.3ex\hbox{$<$}\kern-0.75em{\lower0.65ex\hbox{$\sim$}}}
\def\gsim{\raise0.3ex\hbox{$>$}\kern-0.75em{\lower0.65ex\hbox{$\sim$}}}

\title{MHD Simulations of a Moving Subclump with Heat Conduction}

\author{Naoki Asai$^{1}$, Naoya Fukuda$^{2}$, and Ryoji Matsumoto$^{3}$}
\address{$^{1}$ Graduate School of Science and Technology, Chiba University, 
1-33 Yayoi-cho, Inage-ku, Chiba 263-8522, Japan\\
{\it E-mail: asai@astro.s.chiba-u.ac.jp}}
\address{$^{2}$ Department of Computer Simulation, Faculty of Informatics, 
Okayama University of Science, 1-1 Ridai-cho, Okayama 700-0005, Japan\\
{\it E-mail: fukudany@sp.ous.ac.jp}}
\address{$^{3}$ Department of Physics, Faculty of Science, Chiba University, 
1-33 Yayoi-cho, Inage-ku, Chiba 263-8522, Japan\\
{\it E-mail: matumoto@astro.s.chiba-u.ac.jp}}

%\address{\normalsize{\it (Received October 31, 2004; Accepted December 1,2004)}}

\abstract{
High resolution observations of cluster of galaxies by {\it Chandra} have 
revealed the existence of an X-ray emitting {\it comet-like} galaxy C153 in 
the core of cluster of galaxies A2125. The galaxy C153 moving fast
in the cluster core has a distinct X-ray tail on one side, obviously due to ram 
pressure stripping, since the galaxy C153 crossed the central region of A2125.
The X-ray emitting plasma in the tail is substantially cooler than the ambient plasma.
We present results of two-dimensional magnetohydrodynamic simulations of 
the time evolution of a subclump like C153 moving in magnetized intergalactic matter.
Anisotropic heat conduction is included.
We found that the magnetic fields are essential for the existence of the cool X-ray tail, 
because in non-magnetized plasma the cooler subclump tail is heated up by 
isotropic heat conduction from the hot ambient plasma and does not form such a {\it comet-like} tail.
}


\maketitle

\section {Introduction}

High spatial resolution observations of central region of cluster of galaxies 
by {\it Chandra} have revealed the existence of a peculiar X-ray emitting {\it comet-like} structure.

Wang \etal (2004) presented results of a deep {\it Chandra} observation 
together with extensive multi-wavelength data of 
large-scale hierarchical structure related with A2125. 
An interesting feature is a distinct X-ray tail on one side of 
the fast moving ($v \sim 1500 \, {\rm km \, s^{-1}}$) galaxy C153, 
probably created by ram pressure stripping. 
They suggested that C153 crossed the central region of A2125 containing cD-like 
elliptical galaxies quite recently. 
Since X-ray emission above $1.5 \,{\rm keV}$  is absent in this tail,
this tail is substantially cooler $(kT \leq 1.5 \, {\rm keV})$ than 
ambient intergalactic plasmas $(kT \sim 3.2 \, {\rm keV})$. 
They estimated that the length of this trail is 
$\sim 22^{\prime \prime} ( \sim 88 \, {\rm kpc})$ and its average width is 
$\sim 4^{\prime\prime} (\sim 16 \, {\rm kpc})$. 
Additionally, an extended [O II] line emission 
toward the same direction has been detected.

Since heat conduction in cluster of galaxies can be 
very efficient (e.g., Takahara \& Ikeuchi 1977), heat conduction 
plays a key role in the formation of cold fronts (e.g., Ettori \& Fabian 2000; Vikhlinin \etal 2001) 
and thermal balance in the cluster core (cooling flow problem). 
Asai \etal (2004) showed the dramatic effect of magnetic fields on heat conduction 
in cluster of galaxies.
They carried out magnetohydrodynamic (MHD) simulations of a subcluster moving 
in a magnetized intercluster plasma. 
They showed that the contact surface (a cold front) between the cool subcluster plasma and 
hot intercluster plasma is maintained because the heat conduction across the cold front is 
suppressed by magnetic fields wrapping the forehead of the moving subcluster. 
In non-magnetized plasma, however, the cold front disappears by heat conduction.
Similarly, the cool X-ray tail embedded in hotter ambient plasma should subject to heating 
by thermal conduction.

In this paper, we investigate the interaction between a moving subclump 
and magnetized intergalactic plasma and explore necessary conditions for the existence 
of the cool X-ray trail.

\section {Simulation Model}

We simulated the time evolution of a subclump in a 
frame comoving with the subclump.
We solve the two-dimensional (2D) resistive MHD equations in a Cartesian coordinate 
system $(x, \, y)$. 
We use the specific heat ratio $\gamma = 5/3$.
We assume an anomalous resistivity. The resistivity sets in locally only when 
the drift velocity ($v_{\rm d} \equiv |j|/ \rho$) exceeds the critical velocity ($v_{\rm c}$),
where $j$ is current density. When $v_{\rm d} \geq v_{\rm c}$, the resistivity is 
$\eta = \eta_{0}(v_{\rm d} / v_{\rm c}- 1)^{2}$, 
otherwise $\eta = 0$. We adopt $\eta_{0}= 0.01$ and $v_{\rm c} = 3.0$. 
We set an upper limit of the resistivity, $\eta_{\rm max}=1.0$.
We assume heat conduction along the magnetic field line. 
The classical Spitzer conductivity (Spitzer 1962)
is assumed. When magnetic fields exist, the conductivity along the field line is 
$\kappa \approx \kappa_{\parallel} = \kappa_{0} T^{5/2}
(\kappa_{0} = 5 \times 10^{-7} \, {\rm ergs \, s^{-1} 
\, cm^{-1} \, K^{-1}}$) and the conductivity across the field line is $ \kappa_{\perp} = 0$, 
where $T$ is temperature.

The size of the computational box is $600 \, {\rm kpc} \times 400 \, {\rm kpc}$ 
and the number of grid points is $(N_{x}, \, N_{y})=(1200, 800)$.

The units of length, velocity, density, pressure, temperature, and time
in our simulations are
$r_{0}=25 \, {\rm kpc}$,
$v_{0}=500 \, {\rm km \, s^{-1}}$,
$\rho_{0}=6.5 \times 10^{-27} \, {\rm g \, cm^{-3}}$,
$p_{0} = 4 \times 10^{-11} \, {\rm erg \, cm^{-3}}$,
$T_{0}= 1.5 \, {\rm keV}$, and
$t_{{0}}=r_{0}/v_{0}= 5 \times 10^{7} \, {\rm yr}$,
respectively.

\begin {figure}[h]
\vskip 0cm
\centerline{\epsfxsize=6.cm\epsfbox{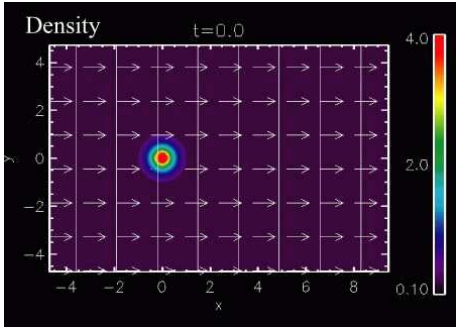}}
\vskip -0.2cm
\label{fig2}
\caption{
Initial density distribution (color scale) in the central region
($375 \, {\rm kpc} \times 250 \, {\rm kpc}$) of simulation box.
Solid lines show magnetic field lines and arrows show velocity vectors.
}
\end{figure}

Figure 1 shows the initial density distribution in the central region 
($375 \, {\rm kpc} \times 250 \, {\rm kpc}$).
Solid lines and arrows show magnetic field lines and velocity vectors.
The subclump is assumed to be a spherical isothermal low-temperature
($kT_{\rm in} = 1.5 \, {\rm keV}$) plasma confined by the gravitational
potential of the subclump. It is embedded in the 
less-dense $(\rho_{\rm out} = 0.25 \, \rho_{0})$, uniform hot 
$(kT_{\rm out} = 3.0 \, {\rm keV})$ plasma. Here the subscripts ``in'' 
and ``out'' denote the values inside and outside the subclump, respectively.
We assume that the density distribution of the subclump is given by the $\beta$-model profile,
$\rho_{\rm in}=\rho_{\rm c} [1 +(r/r_{\rm c})^{2}]^{-3 \beta/2}$,
where $r = (x^{2}+y^{2})^{1/2}$ and $\beta=2/3$. The maximum density is 
$\rho_{\rm c}=5 \rho_{0} = 3.2 \times 10^{-26} \, {\rm g \, cm^{-3}}$, and 
the core radius is $r_{\rm c} = 8.3 \, {\rm kpc}$. 
The subclump is initially in hydrostatic equilibrium under the gravitational 
potential fixed throughout the simulation.

Table 1 shows model parameters.
Important parameter is the plasma beta ($\beta_{0}$) defined as the ratio of 
the ambient gas pressure to the magnetic pressure.
When $\beta_{0} = p_{\rm gas} / p_{\rm mag}= 100$, the magnetic field strength is
$B \sim 3 \, {\rm \mu G}$. 

We assume that the ambient plasma initially has a uniform speed with 
Mach number $M = v_{x}/c_{\rm s \, out} = 1$,
where $c_{\rm s \, out}$ is the ambient sound speed. The Mach number 
with respect to the sound velocity inside the subcluster is 
$ M^{\prime}= v_{x}/c_{\rm s \, in}= \sqrt{2}$. 

Model HC is a non-magnetic model with isotropic heat conduction. 
Models MCa, MCb, and MCc are models with moderate magnetic fields 
($\beta_{0}=100$) and anisotropic heat conduction. 
In models HC, MCb, and MCc, the Mach number is taken to be $M=2$, and in model MCa, 
it is taken to be $M=1$.
The inclination of magnetic fields from motion of the subclump is 
parameterized by $\theta=\arccos [{\bf v \cdot B }/(v \, B)]$.

\begin{table}
\begin{center}
\caption{Models and parameters. 
Heat conduction is isotropic in model HC. 
In other models, heat conducts only along magnetic fields.
$\theta$ is an angle between the motion of the subclump and the magnetic field.
}
\label{tbl-2}
\begin{tabular}[t]{ccccc}
\hline
\hline
Model & $\theta [{}^{\circ}]$ & $\beta_{0}$  & $\kappa$ & Mach number  \\
\hline
MCa & 90  &100 & $\kappa_{\parallel}$ & 1 \\
MCb & 90  &100 & $\kappa_{\parallel}$ & 2 \\
MCc & 45  &100 & $\kappa_{\parallel}$ & 2 \\
HC &  --- &$\infty$ & $\kappa$  & 2 \\
\hline
\end{tabular}
\end{center}
\end{table}

We use a modified Lax-Wendroff method with artificial viscosity
for MHD part, and the heat conduction term in energy equation is solved by 
the implicit red and black successive overrelaxation method 
(see Yokoyama \& Shibata 2001 for details). 

For boundary conditions, the left boundary at $x= -5$ is 
taken to be a fixed boundary, and other boundaries are taken to be
free boundaries where waves can be transmitted.

\section {Numerical Results}\label{res}
\subsection {Effects of Magnetic Fields on Cold X-ray Tail}\label{tail}

\begin {figure*}[t]
\vskip 0cm
\centerline{\epsfxsize=17cm\epsfbox{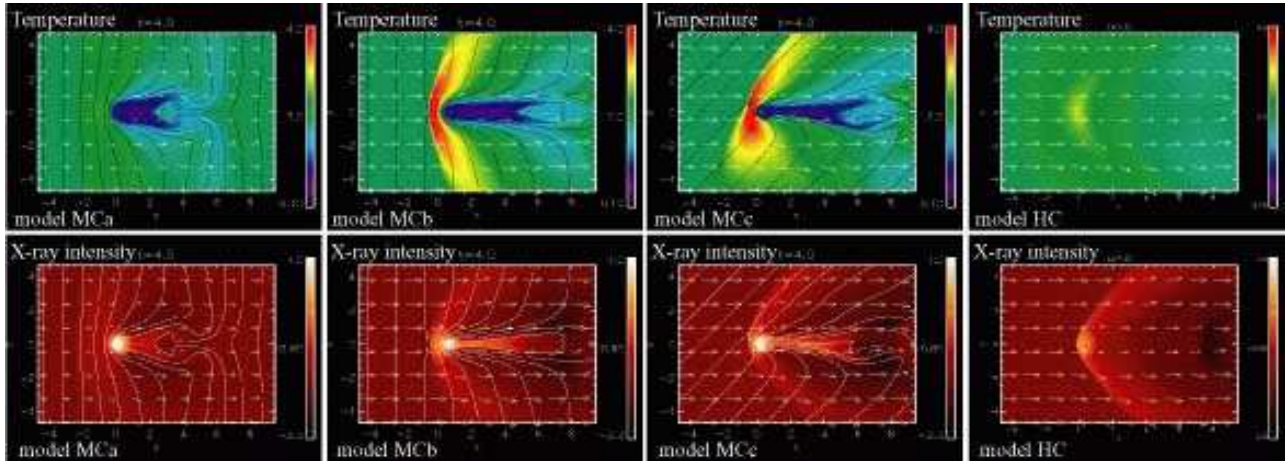}}
\vskip -0.2cm
\label{fig2}
\caption{Results for models MCa, MCb, MCc, and HC from left to right, respectively. 
The top and bottom panels show distribution of temperature ({\it top}) and 
X-ray intensity ({\it bottom}) at the central region ($375 \, {\rm kpc} \times 250 \, {\rm kpc}$) 
after $t= 2 \times 10^{8} \, {\rm yr}$, respectively.
Solid curves show the magnetic field lines and arrows show the velocity vectors.}
\vskip -0.5cm
\end{figure*}

Let us investigate effects of magnetic fields on the existence of the cold X-ray tail.
Figure 2 shows the results for models MCa, MCb, MCc, and HC from left to right, respectively. 
The top and bottom panels in Figure 2 show the distributions of 
temperature and X-ray intensity at the central region ($375 \, {\rm kpc} \times 250 \,
{\rm kpc}$) after $t= 2 \times 10^{8} \, {\rm yr}$. X-ray intensity is 
visualized from simulation results as logarithm of $\sim \rho^{2}$.

In models with magnetic fields (MCa, MCb, and MCc), the cold plasma inside the subclump 
is expelled backward due to ram pressure and forms a cool tail dividing 
into two branches. The cool tail survives the heat conduction from hotter 
ambient plasmas, because magnetic fields wrapping the subclump
suppress the heat conduction across them. 
On the other hand, in a model without magnetic fields (model HC),
cool plasmas inside the subclump is heated by thermal conduction and evaporates quickly.
Thus cool tail is not formed in this model.

In all models with magnetic fields, magnetic fields accumulating ahead of the subclump 
form a magnetic shield (e.g., Miniati \etal 1999) and 
their strength is enhanced several times that of the initial state.
In addition to the suppression of heat conduction,
magnetic fields also prevent the Kelvin-Helmholtz instability in this region.

A typical length of the X-ray tail formed in our simulation is $100-200 \,{\rm kpc}$, 
and its width is $25-50 \,{\rm kpc}$ at $t=2 \times 10^{8} {\rm yr}$.
The tails are longer in models MCb and MCc (Mach number $M=2$), than model MCa (Mach number $M=1$),  
because the subclump subjects to the stronger ram pressure.

\subsection {Effects of Magnetic Fields on Energy Conversion}\label{con}

When a subclump moves in magnetized plasmas, 
magnetic fields can extract the kinetic energy of the plasma moving with the subclump
because magnetic field lines are stretched by the plasma moving with the sublcump.
The kinetic energy is converted to magnetic energy and thermal energy.
We computed the energy conversion rates in order to study whether a motion of a subclump 
heats the ambient plasma.
Figure 3 shows the time evolution of magnetic ({\it solid line}), kinetic ({\it dotted line}), 
and thermal ({\it dot-dashed line}) energies with respect to the initial kinetic energy, 
respectively. The left and right panels show results 
for model MCb (with magnetic fields) and HC (without magnetic fields).

The left panel shows that magnetic energy increases only slightly, 
because magnetic fields are deformed in small area close to the subclump.
On the other hand, thermal energy increases while kinetic energy decreases.
That is, the kinetic energy of the subclump is converted into thermal energy through
a shock heating. The right panel shows the similar behavior in model HC.

The inefficiency of the energy conversion from kinetic energy to magnetic energy 
is partly due to the free boundary condition. 
When the magnetic fields are fixed at the boundaries,
magnetic fields will be deformed until magnetic energy is comparable to 
the gravitational energy.

% Therefore, for a plasma heating, effect of magnetic process on energy conversion into 
% thermal energy is not prominent in this local simulation. 

\begin {figure}[h]
\vskip 0cm
\centerline{\epsfxsize=8.5cm\epsfbox{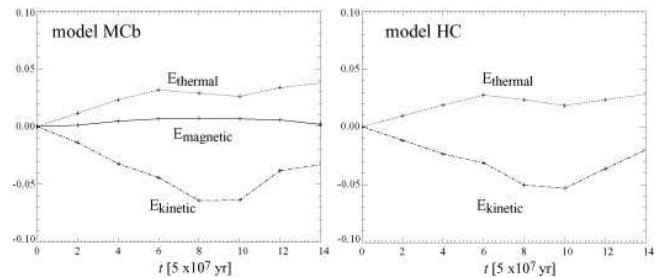}}
\vskip -0.2cm
\label{fig3}
\caption{Time evolutions of magnetic ({\it solid line}), kinetic ({\it dot-dashed line}),
and thermal ({\it dotted line}) energies integrated for the whole simulation region, respectively. 
Left: Results for model MCb (with magnetic fields). 
Right: Results for model HC (without magnetic fields).}
\vskip -0.5cm
\end{figure}

\section {Discussion \& Summary}\label{dis}

We carried out 2D MHD simulations of a subclump moving through 
a magnetized ambient plasma. 
In \S \ref{res}\ref{tail}, we showed that the magnetic field is essential for
the existence of a cold X-ray tail of a subclump like the galaxy C153 observed in A2125.
Heat conduction across the magnetic fields is suppressed by the magnetic field 
wrapping the subclump. This mechanism is the same as that which enables the maintenance
of cold fronts in cluster of galaxies (Asai \etal 2004).

In the context of interstellar matter, similar works have been done 
(e.g., Jones \etal 1996; Miniati \etal 1999), 
although heat conduction and gravity are not included in their simulations.

They investigated, through 2D MHD simulations, the interaction of 
a uniform magnetic field oblique to a moving interstellar cloud.
Miniati \etal (1999) discussed the conversion rate of kinetic energy
to magnetic energy for several models. 
In contrast to their results, only a small fraction of the kinetic energy is 
converted to the magnetic energy in our models (see \S \ref{res}\ref{con}). 
The difference comes mainly from the fact that a subclump in our model has 
lower density and moves faster than that in their model.

Makishima \etal (2001)  proposed a model of heating of cluster plasma through the 
motion of member galaxies in magnetic fields. In local simulations we presented
in this papar, the energy conversion rate of kinetic energy of the moving subclump 
to the magnetic energy is small because magnetic fields can freely move at boundaries.
In cluster of galaxies, magnetic fields may be anchored to the cD galaxies. When 
subclumps move in such magnetosphere of the cD galaxy, magnetic fields will be 
stretched and twisted. Under such situation, the kinetic energy of the dark matter clump 
will be extracted through the magnetic interaction.
The deformed magnetic field lines may form current sheets, in which 
magnetic reconnection converts magnetic energy into thermal energy and kinetic energy.
We expect efficient heating through this process. This mechanism can be the 
heat source which compensates for the radiative cooling in cluster plasmas.
Obviously, we have to carry out global MHD simulations to study this process. We would 
like to report the results of such simulations 
in subsequent papers (Asai \etal 2005 in preparation).

\acknowledgements{
We thank T. Yokoyama for developments of the coordinated astronomical
numerical software (CANS) which include 2D MHD codes including heat conduction.
The development of CANS was supported by ACT-JST of Japan Science and Technology Corporation.
This work is supported by the priority research project in 
graduate school of Science and Technology, Chiba University (P.I., R. Matsumoto).
Numerical computations are carried out by VPP5000 at NAOJ.
}

\end{document}